\documentclass[12pt,epsf,fleqn]{article}
\usepackage{epsfig}
\usepackage[dvips]{color}
\begin{document}
\title{The Higgs search of the MSSM with explicit CP violation at the LHC and ILC}
\author{S.W. Ham$^{(1)}$, S.G. Jo$^{(2)}$, S.K. Oh$^{(1,3)}$, and D. Son$^{(1)}$
\\
\\
{\it $^{\rm (1)}$ Center for High Energy Physics, Kyungpook National University}\\
{\it Daegu 702-701, Korea} \\
{\it $^{\rm (2)}$ Department of Physics, Kyungpook National University} \\
{\it Daegu 702-701, Korea} \\
{\it $^{\rm (3)}$ Department of Physics, Konkuk University, Seoul
143-701, Korea}
\\
\\
}
\date{}
\maketitle
\begin{abstract}
We study the neutral Higgs sector of the minimal supersymmetric standard model (MSSM)
with explicit CP violation at the one-loop level.
We take into account the one-loop contributions by the top quark, the stop quarks, the bottom quark,
the sbottom quarks, the tau lepton, the stau leptons, the $W$ boson, the charged Higgs boson,
the charginos, the $Z$ boson, the neutral Higgs bosons, and the neutralinos.
The production cross sections of the neutral Higgs boson are calculated to the leading order.
The processes in our consideration are divided in two groups: the Higgs-strahlung and gluon fusion processes
accessible at the CERN Large Hadron Collider (LHC), and the vector boson fusion and Higgs-strahlung
processes accessible at the $e^+e^-$ International Linear Collider (ILC).
In particular, we investigate the dependence of these processes on the CP phase arising
from the $U(1)$ factor of the gaugino mass in the neutralino mass matrix.
We show that the cross sections of these processes vary by the range of 3\% $-$ 19 \%
as the CP phase changes from zero to $\pi$.
\end{abstract}
\vfil \eject

\section{Introduction}

The violation of CP symmetry has been observed in the neutral kaon system more than four decades ago [1].
In the Standard Model (SM), the CP violation can be induced by a complex phase
in the Cabibbo-Kobayashi-Maskawa matrix for the charged weak current [2].
Alternatively, it is known that if a model possesses at least two Higgs doublets,
the CP violation may occur in its Higgs sector through the mixing
between the CP even and the CP odd states [3].
It is one of the characteristics of the minimal supersymmetric standard model (MSSM)
that it should possess two Higgs doublets in order to give masses
to the up-quark sector and the down-quark sector separately [4-7].
Thus, in principle, the MSSM may accommodate the CP violation by means of complex phases
in its neutral Higgs sector.
In practice, it has been found that the CP violation is impossible to occur
either explicitly or spontaneously in the Higgs sector of the MSSM at the tree level,
because the complex phases can always be eliminated by rotating the Higgs fields.

The possibility of CP violation in the MSSM at the one-loop level has been studied by many authors [8-38].
It has been noticed that the spontaneous CP violation in the Higgs sector of the MSSM
at the one-loop level is disfavored because it requires a very light neutral Higgs boson,
which has already been ruled out by experiments.
The explicit CP violation, on the other hand, is viable by virtue of the radiative corrections
due to the loops of relevant particles such as the quarks, the squarks, the $W$ boson,
the charged Higgs boson, the charginos, the $Z$ boson, the neutral Higgs bosons,
and the neutralinos.
The radiative corrections by these particles yield the mixing
between the CP even and the CP odd neutral Higgs bosons.
Thus, it is possible to achieve the explicit CP violation in the radiatively corrected Higgs sector of the MSSM.

Recently, a number of studies have been devoted to the prospects for discovering neutral Higgs bosons
in a general two Higgs doublet model (THDM) [39-41]
and in the MSSM [34-36] in high-energy $e^+e^-$ collisions
within the context of explicit CP violation.
In the MSSM with explicit CP violation, where the CP mixing among the CP even
and the CP odd states is maximized, the cross sections for the neutral Higgs boson productions
in $e^+ e^-$ collisions at $\sqrt{s}$ = 500 and 800 GeV have been calculated [36].
Also, in the context of the explicit CP violation scenario the production
cross sections of the neutral Higgs boson in hadron colliders have been calculated
by considering the Higgs-strahlung process which is associated with the weak gauge bosons [37]
and the gluon fusion process [38].

In this article, we investigate the phenomenological implication of the CP phase arising
from the neutralino contribution which contributes to the CP mixing
among the scalar and pseudoscalar Higgs bosons on the Higgs search at the future high-energy colliders,
such as the CERN Large Hadron Collider (LHC) and the $e^+e^-$ International Linear Collider (ILC).
Within the framework of the explicit CP violation in the MSSM at the one-loop level,
we calculate the production cross sections of three neutral Higgs bosons
at the leading order in high-energy $e^+e^-$ and $PP$ collisions.
At the LHC, the three dominant Higgs productions are considered to be the gluon fusion process and
the associated Higgs-strahlung processes with the weak gauge bosons($Z$, $W$).
At the ILC, the three dominant mechanisms for the neutral Higgs productions are considered
to be the vector boson ($Z$, $W$) fusion processes and the Higgs-strahlung process.
We pay attention to the dependence of the Higgs production on the CP phase
whose appearance comes from the neutralino contribution to the tree-level Higgs sector.
In the MSSM with explicit CP violation, the production cross sections of the neutral Higgs bosons
at both LHC and ILC are shown to vary about 3\%$-$19 \%
with respect to the variation of the CP phase, arising from the $U(1)$ factor of the gaugino mass parameter,
from zero to $\pi$.

\section{Higgs sector}

The Higgs sector of the MSSM consists of two Higgs
doublets, $H_1$ and $H_2$. In terms of these Higgs doublets, the
superpotential of the MSSM is given by
\begin{equation}
    W = h_b Q b_R^c H_1 + h_t Q t_R^c H_2 + h_{\tau} L H_1 \tau_R^c - \mu H_1 H_2 \ ,
\end{equation}
where we take into account only the third generation: $Q$ and $L$
are the SU(2) doublet quark and lepton superfields of the third
generation respectively, $t_R^c$, $b_R^c$ and $\tau_R^c$ are the
SU(2) singlet top, bottom, and tau superfields respectively, $h_t$,
$h_b$, $h_{\tau}$ are the Yukawa coupling coefficients of top,
bottom, and tau superfields respectively, and $\mu$ is the Higgs
mixing parameter with mass dimension.

The Higgs potential at the tree level reads
\begin{eqnarray}
    V^0 & = & {g_2^2\over 8} (H_1^{\dag} \vec\sigma H_1+ H_2^{\dag} \vec\sigma H_2)^2 +
    {g_1^2\over 8}(|H_2|^2 - |H_1|^2)^2   \cr
    & &\mbox{} + m_1^2 |H_1|^2 + m_2^2 |H_2|^2 - m_3^2 (H_1^T \epsilon H_2 + {\rm H.c.}) \ ,
\end{eqnarray}
where $\epsilon$ is an antisymmetric $2 \times 2$  matrix with $\epsilon_{12} = 1$,
$\vec\sigma$ denotes the three Pauli matrices,
$g_1$ and $g_2$ are the U(1) and SU(2) gauge coupling constants respectively,
and $m_i^2$ $(i = 1, 2, 3)$ are the soft SUSY breaking masses,
which may be assumed to be real without loss of generality.
We choose $m_3^2$ to be positive.
Using minimum conditions with respect to the neutral Higgs fields,
we may eliminate $m^2_1$ and $m^2_2$.
Thus, only $m^2_3$ remains as a free parameter.

The two Higgs doublets may be expressed as
\begin{eqnarray}
\begin{array}{lll}
            H_1 & = &
    \displaystyle{{1 \over \sqrt{2}}} \left ( \begin{array}{c}
            v_1 + h_1 + i h_3 \sin \beta   \cr
            C^{+ *} \sin \beta
    \end{array} \right )  \ ,  \cr
            H_2 & = &
    \displaystyle{{1 \over \sqrt{2}}} \left ( \begin{array}{c}
            C^+ \cos \beta           \cr
            v_2 + h_2 + i h_3 \cos \beta
    \end{array} \right ) e^{i \phi_0} \ ,
\end{array}
\end{eqnarray}
where $v_1$ and $v_2$ are the vacuum expectation values (VEVs) of the neutral Higgs fields,
$\tan\beta = v_2 / v_1$, and $\phi_0$ is the relative phase between the two Higgs doublets.
The five physical Higgs fields are three real neutral Higgs fields $h_1$, $h_2$, $h_3$
and one complex charged Higgs field $C^+$ carrying two real degrees of freedom.

The tree-level masses of the fermions of the third generation are given as $m_t = h_t v \sin\beta/\sqrt{2}$,
$m_b = h_b v \cos\beta/\sqrt{2}$, $m_{\tau} = h_{\tau} v \cos\beta/\sqrt{2}$,
and the tree-level masses of the weak gauge boson are given as $m_W^2 = g_2^2 v^2 /4$
and $m_Z^2 = (g_1^2 + g_2^2) v^2 /4$, where $v = \sqrt{v^2_1 + v^2_2} = 246$ GeV.
In terms of these masses, we can express the masses of the Higgs bosons and those of superpartners.

At the tree level, $\phi_0$ may be taken to be zero.
Thus, at the tree level, CP violation can be avoided in the MSSM and the three neutral Higgs bosons
are divided into the CP even states $h^0$, $H^0$, and the CP odd state $A^0$.
The tree-level mass of $A^0$ is
\[
    m_A^2 = {2 m_3^2 \cos \phi_0 \over \sin 2 \beta}  \ ,
\]
while the tree-level masses of $h^0$ and $H^0$ are
\[
    m^2_h, m^2_H  =
    {1 \over 2} \left [m_Z^2 + m_A^2 \mp
    \sqrt{ \left (m_Z^2 + m_A^2 \right )^2 - 4 m_Z^2 m_A^2 \cos^2 2 \beta}  \right ]     \ ,
\]
where $m^2_h \le m^2_H$ is understood. The tree-level mass of the
charged Higgs boson $C^+$ is
\[
    m_C^2 = m_W^2 + m_A^2 \ .
\]

For the masses of superpartners, the tree-level masses of the scalar
fermions of the third generation are
\begin{eqnarray}
    m_{{\tilde t}_1}^2, m_{{\tilde t}_2}^2 & = &
    m_t^2 + {1 \over 2}(m_Q^2 + m_T^2) + {m_Z^2 \over 4} \cos 2 \beta  \mp \sqrt{X_{\tilde t}}  \ , \cr
    m_{{\tilde b}_1}^2, m_{{\tilde b}_2}^2 & = &
    m_b^2 + {1 \over 2} (m_Q^2 + m_B^2) - {m_Z^2 \over 4} \cos 2 \beta \mp \sqrt{X_{\tilde b}}  \ , \cr
    m_{{\tilde \tau}_1}^2, m_{{\tilde \tau}_2}^2 & = &
    m_{\tau}^2 + {1 \over 2}(m_L^2 + m_E^2) - {m_Z^2 \over 4} \cos 2 \beta \mp \sqrt{X_{\tilde \tau}}  \ ,
\end{eqnarray}
with
\begin{eqnarray}
X_{\tilde t} & = & \left \{ {1 \over 2} (m_Q^2 - m_T^2)
    + \left ( {2 \over 3} m_W^2 - {5 \over 12} m_Z^2 \right ) \cos 2 \beta \right \}^2 \cr
& &\mbox{}        + m_t^2 (A_t^2  + \mu^2 \cot^2 \beta - 2 A_t \mu
\cot \beta \cos \phi_t)     \ , \cr X_{\tilde b} & = & \left \{ {1
\over 2} (m_Q^2 - m_B^2)
    + \left ( {1 \over 12} m_Z^2 - {1 \over 3} m_W^2 \right ) \cos 2 \beta \right \}^2 \cr
& &\mbox{}        + m_b^2 (A_b^2 + \mu^2 \tan^2 \beta - 2 A_b \mu
\tan \beta \cos \phi_b)   \ , \cr X_{\tilde \tau} & = & \left \{ {1
\over 2} (m_L^2 - m_E^2) + \left ( {3 \over 8} m_Z^2 - {1 \over 2}
m_W^2 \right ) \cos 2 \beta \right \}^2 \cr & &\mbox{}        +
m_{\tau}^2 (A_{\tau}^2 + \mu^2 \tan^2 \beta  - 2 A_{\tau} \mu \tan
\beta \cos \phi_{\tau})    \ ,
\end{eqnarray}
where $A_t$, $A_b$, and $A_{\tau}$ are trilinear soft SUSY breaking parameters
coming from the scalar quark and lepton sectors of the third generation,
and $m_T$, $m_B$, and $m_Q$ are the quark singlets and doublet soft masses, and $m_E$ and $m_L$
are the lepton singlet and doublet soft masses.
The tree-level masses of the charginos are
\begin{eqnarray}
    m_{{\tilde \chi}_1}^2 \ , m_{{\tilde \chi}_2}^2 & = &
    {1 \over 2} (M_2^2 + \mu^2) + m_W^2 \mp \sqrt{X_{\tilde \chi}} \ ,
\end{eqnarray}
with
\begin{eqnarray}
X_{\tilde \chi} & = & \left \{ {1 \over 2} (M_2^2 - \mu^2) - m_W^2
\cos 2 \beta \right \}^2 \cr
    & &\mbox{} + 2 m_W^2 \cos^2 \beta (M_2^2 + \mu^2 \tan^2 \beta + 2 M_2 \mu \tan \beta \cos \phi_c)  \
    .
\end{eqnarray}
Note that the CP violating phases $\phi_t$, $\phi_b$, $\phi_{\tau}$, and $\phi_c$,
stemming from the stop quark, the sbottom quark, the stau lepton,
and the chargino contributions, appear in the above expression.
Finally, the tree-level masses of the neutralinos are given as the eigenvalues of the neutralino mass matrix
\[
        {\cal M}_{{\tilde \chi}^0} =
    \left (
    \displaystyle{
    \begin{array}{cccc}
    M_1 e^{i\phi_1} & 0 & -g_1 v_1 /2 & g_1 v_2 /2  \cr
     & & & \cr
    0 & M_2 & g_2 v_1 /2 & -g_2 v_2 /2  \cr
    & & & \cr
    -g_1 v_1 /2 & g_2 v_1 /2 & 0 & -\mu e^{i\phi_2}  \cr
    & & & \cr
    g_1 v_2 /2 & -g_2 v_2 /2 & -\mu e^{i\phi_2} & 0
\end{array}  }
\right )  \ .
\]
Here, too, two CP violating phases $\phi_1$ and $\phi_2$ are present.
The CP phase $\phi_1$ is the relative phase between $M_1$ and $M_2$,
and $\phi_2$ is the relative phase between $M_2$ and $\mu$.
Thus, $\phi_2$ is identical to $\phi_c$ and the additional CP phase
arising from the neutralino sector is only $\phi_1$.

At the one-loop level, the explicit CP violation would occur and the neutral Higgs bosons
can no longer have definite CP eigenvalues.
The CP even states and the CP odd state are mixed to yield $h_i$ ($i$ = 1,2,3).
We employ the effective potential method in order to calculate radiative corrections [42].
The one-loop effective potential is given as
\[
    V^1 =
    \sum_k {c_k \over 64 \pi^2} (-1)^{2 J_k} (2 J_k + 1) {\cal M}^4_k (h_i)
    \left [\log \left ( { {\cal M}^2_k (h_i) \over \Lambda^2} \right ) - {3\over 2} \right ]  \ ,
\]
where the summation over $k$ is carried out for all contributions, namely,
the contributions of the top quark, the stop quarks, bottom quark,
the sbottom quarks, the tau lepton, the stau leptons, the $W$ boson, the charged Higgs boson,
the charginos, the $Z$ boson, the neutral Higgs bosons, and the neutralinos;
and $J_k$ is the spin of the corresponding particle, $\Lambda$ is the renormalization scale in the modified
minimal subtraction $(\overline {\rm MS})$ scheme,
and $c_k$ is the color factor $c_{\rm colour}$ multiplied by the charge factor $c_{\rm charge}$.
The color factors for colored and uncolored particles are 3 and 1 respectively,
and the charge factors for charged and neutral particles are 2 and 1 respectively.
Note that in this one-loop effective potential ${\cal M}_k^2$ depends
on the neutral Higgs fields $h_i$ ($i$ = 1, 2, 3), and
it contains the tree-level masses of relevant particles and superpartners.

There is a CP-odd tadpole minimum condition with respect to $h_3$ at
the one-loop level in the MSSM, which is given as
\begin{eqnarray}
    0 & = &
    m_3^2 \sin \phi_0 +
        {3 m_t^2 \mu A_t \sin \phi_t \over 16 \pi^2 v^2 \sin^2 \beta} f_1 (m_{{\tilde t}_1}^2,  \ m_{{\tilde t}_2}^2)
    + {3 m_b^2 \mu A_b \sin \phi_b \over 16 \pi^2 v^2 \cos^2 \beta}
    f_1 (m_{{\tilde b}_1}^2,  \ m_{{\tilde b}_2}^2) \cr
    & &\mbox{}
    + {m_{\tau}^2 \mu A_{\tau} \sin \phi_{\tau} \over 16 \pi^2 v^2 \cos^2 \beta}
        f_1 (m_{{\tilde \tau}_1}^2,  \ m_{{\tilde \tau}_2}^2)
    -  {m_W^2 \mu M_2 \sin \phi_c \over 4  \pi^2 v^2} f_1 (m_{{\tilde  \chi}_1}^2, \ m_{{\tilde \chi}_2}^2) \cr
    & &\mbox{}
    + \sum^4_{k = 1} {m_{{\tilde \chi}^0_k}^2  \over 4 \pi^2 v^2}
    \left (\log{m_{{\tilde \chi}^0_k}^2 \over \Lambda^2} - 1 \right)
    {E (m_{{\tilde \chi}^0_k}^2 - \mu^2 ) \mu
    \over \prod\limits_{a \not= k} (m_{{\tilde \chi}^0_k}^2 - m_{{\tilde
    \chi}^0_a}^2)}\, ,
\end{eqnarray}
where the six terms on the right-hand side, in the order of the appearance,
come from the tree-level Higgs potential, the one-loop contributions of the stop quark,
the sbottom quark, the stau lepton, the chargino, and the neutralinos, and
\[
        E = -\left[ (m_{{\tilde \chi}^0_k}^2 - M_1^2) M_2  m_W^2 \sin \phi_2
    + (m_{{\tilde \chi}^0_k}^2 - M_2^2)  M_1 (m_Z^2 - m_W^2) \sin(\phi_1 + \phi_2)
    \right] \ ,
\]
and the dimensionless function $f_1(m_x^2, m_y^2)$ is defined as
\[
    f_1 (m_x^2, \  m_y^2) =
    {2 \over (m_x^2-m_y^2)} \left \{m_x^2 \log  {m_x^2 \over
            \Lambda^2} -m_y^2 \log {m_y^2 \over \Lambda^2} \right \} - 2 \ .
\]

At the one-loop level, the matrix elements of $M_{ij}$ in the ($h_1,h_2,h_3$) basis are given by
\begin{equation}
M_h = \left ( \begin{array}{ccc}
    M_{11} & M_{12} & M_{13}   \cr
    M_{12} & M_{22} & M_{23}   \cr
    M_{13} & M_{23} & M_{33}
        \end{array}
    \right ) \ ,
\end{equation}
where
\begin{eqnarray}
M_{11} & = & m_Z^2 \cos^2 \beta + {\bar m}_A^2 \sin^2 \beta +
M_{11}^1 \ ,  \cr M_{22} & = & m_Z^2 \sin^2 \beta + {\bar m}_A^2
\cos^2 \beta + M_{22}^1 \ ,  \cr M_{33} & = & {\bar m}_A^2 +
M_{33}^1 \ , \cr M_{12} & = &\mbox{} - (m_Z^2 + {\bar m}_A^2) \cos
\beta \sin \beta + M_{12}^1 \ ,  \cr M_{13} & = & M_{13}^1 \ , \cr
M_{23} & = & M_{23}^1 \ .
\end{eqnarray}
Here the mass parameter ${\bar m}_A$ is defined as
\begin{eqnarray}
{\bar m}_A^2 & = & {2 \over \sin 2 \beta} \left [\rule{0mm}{9mm}
m_3^2 \cos (\phi_0) + {3 m_t^2 \mu A_t \cos \phi_t \over 16 \pi^2
v^2 \sin^2 \beta} f_1 (m_{{\tilde t}_1}^2,  \ m_{{\tilde t}_2}^2)
\right. \cr & &\mbox{}\left. + {3 m_b^2 \mu A_b \cos \phi_b \over 16
\pi^2 v^2 \cos^2 \beta} f_1 (m_{{\tilde b}_1}^2,  \ m_{{\tilde
b}_2}^2)
 + {m_{\tau}^2 \mu A_{\tau} \cos \phi_{\tau} \over 16 \pi^2 v^2 \cos^2 \beta}
f_1 (m_{{\tilde \tau}_1}^2,  \ m_{{\tilde \tau}_2}^2)  \right. \cr &
&\mbox{}\left. +  {m_W^2 \mu M_2 \cos \phi_c \over 4  \pi^2 v^2} f_1
(m_{{\tilde  \chi}_1}^2, \ m_{{\tilde \chi}_2}^2) \right. \cr &
&\mbox{}\left. + \sum^4_{k = 1} {m_{{\tilde \chi}^0_k}^2  \over 4
\pi^2 v^2} \left \{\log \left ({m_{{\tilde \chi}^0_k}^2 \over
\Lambda^2} \right ) - 1 \right \} {E_{h} \over \prod\limits_{a \not=
k} (m_{{\tilde \chi}^0_k}^2 - m_{{\tilde \chi}^0_a}^2)}  \right ] \
,
\end{eqnarray}
with
\begin{eqnarray}
E_{h} & = & (m_{{\tilde \chi}^0_k}^2 - M_1^2) (m_{{\tilde
\chi}^0_k}^2 - \mu^2 ) M_2 \mu m_W^2 \cos \phi_2 \cr & &\mbox{} +
(m_{{\tilde \chi}^0_k}^2 - M_2^2) (m_{{\tilde \chi}^0_k}^2 - \mu^2 )
M_1 \mu (m_Z^2 - m_W^2) \cos(\phi_1 + \phi_2) \ .
\end{eqnarray}
In $M_h$, the matrix elements $M_{i3} = M_{i3}^1$ $(i = 1, 2)$ represent the mixing between the scalar and the
pseudoscalar components.
Thus, nonvanishing of $M_{i3} $ indicates that the mixing occurs at the one-loop level.
There is no mixing at the tree level. Explicit expressions of $M_{ij}^1$ are given in Ref. [27].

The squared mass of the lightest neutral Higgs boson, $m_{h_1}^2$, is given
as the smallest eigenvalue of the one-loop squared-mass matrix for the neutral Higgs bosons.
The upper bound on $m_{h_1}^2$ is obtained by noticing that the smallest eigenvalue of a positive
symmetric matrix cannot exceed the smaller eigenvalue of its upper left $2\times2$ submatrix [43].
Thus, the upper bound on $m_{h_1}^2$ is given as
\begin{equation}
    m_{h_1}^2  \le m_{{h_1}, \ {\rm max}}^2  =
    m_Z^2 \cos^2 2 \beta  + \delta m_{\tilde t}^2 + \delta m_{\tilde b}^2 + \delta m_{\tilde \tau}^2
    + \delta m_{\tilde \chi}^2 + \delta m_h^2 + \delta m_{{\tilde \chi}^0}^2  \ ,
\end{equation}
where the first term comes from the tree-level Higgs potential while the other terms
come from the one-loop corrections due to the top quark, the stop quarks,
the bottom quark, the sbottom quarks, tau lepton, the stau leptons,
the $W$ boson, the charged Higgs boson, the charginos, the $Z$ boson, the neutral Higgs bosons,
and the neutralinos. Explicitly, they are given as follows:
\begin{eqnarray}
    \delta m_{\tilde t}^2 & = &
    \mbox{} - {3 \Delta_{\tilde t}^2 \over 16 \pi^2 v^2}
    {f_2 (m_{{\tilde t}_1}^2, \ m_{{\tilde t}_2}^2) \over (m_{{\tilde t}_2}^2 - m_{{\tilde t}_1}^2)^2}
    + {3 m_t^4 \over 4 \pi^2 v^2} \log \left ({m_{{\tilde t}_1}^2  m_{{\tilde t}_2}^2 \over m_t^4} \right ) \cr
    & &\mbox{} + {3 m_Z^2 \cos 2 \beta \over 64 \pi^2 v^2}
    (\cos 2 \beta m_Z^2 + 8 m_t^2) \log \left ({m_{{\tilde t}_1}^2
    m_{{\tilde t}_2}^2 \over \Lambda^4} \right ) \cr
    & &\mbox{} + {3 \cos^2 2 \beta \over 16 \pi^2 v^2}
    \left ({4 m_W^2 \over 3} - {5 m_Z^2 \over 6} \right )^2 f_1
    (m_{{\tilde t}_1}^2, \ m_{{\tilde t}_2}^2) \cr
    & &\mbox{} + {3 \Delta_{\tilde t} \over 16 \pi^2 v^2}
    (4 m_t^2 + \cos 2 \beta m_Z^2)
    {\log (m_{{\tilde t}_2}^2 / m_{{\tilde t}_1}^2) \over (m_{{\tilde t}_2}^2 - m_{{\tilde t}_1}^2)} \  , \cr
& & \cr
    \delta m_{\tilde b}^2 & = &
    \mbox{} - {3 \Delta_{\tilde b}^2 \over 16 \pi^2 v^2}
    {f_2 (m_{{\tilde b}_1}^2, \ m_{{\tilde b}_2}^2) \over (m_{{\tilde b}_2}^2 - m_{{\tilde b}_1}^2)^2}
    + {3 m_b^4 \over 4 \pi^2 v^2} \log \left ({m_{{\tilde b}_1}^2  m_{{\tilde b}_2}^2 \over m_b^4} \right ) \cr
    & &\mbox{} + {3 m_Z^2 \cos 2 \beta \over 64 \pi^2 v^2}
    (\cos 2 \beta m_Z^2 - 8 m_b^2) \log \left ({m_{{\tilde b}_1}^2
    m_{{\tilde b}_2}^2 \over \Lambda^4} \right ) \cr
    & &\mbox{} + {3 \cos^2 2 \beta \over 16 \pi^2 v^2}
    \left ({m_Z^2 \over 6} - {2 m_W^2 \over 3} \right )^2 f_1 (m_{{\tilde b}_1}^2, \ m_{{\tilde b}_2}^2) \cr
    & &\mbox{} + {3 \Delta_{\tilde b} \over 16 \pi^2 v^2}
    (4 m_b^2 - \cos 2 \beta m_Z^2)
    {\log (m_{{\tilde b}_2}^2 / m_{{\tilde b}_1}^2) \over (m_{{\tilde b}_2}^2 - m_{{\tilde b}_1}^2)} \  , \cr
& & \cr
    \delta m_{\tilde \tau}^2 & = &
    \mbox{} - {\Delta_{\tilde \tau}^2 \over 16 \pi^2 v^2}
    {f_2 (m_{{\tilde \tau}_1}^2, \ m_{{\tilde \tau}_2}^2) \over
    (m_{{\tilde \tau}_2}^2 - m_{{\tilde \tau}_1}^2)^2}
    + {m_{\tau}^4 \over 4 \pi^2 v^2}
    \log \left ({m_{{\tilde \tau}_1}^2  m_{{\tilde \tau}_2}^2 \over m_{\tau}^4} \right ) \cr
    & &\mbox{}
    + {m_Z^2 \cos 2 \beta \over 64 \pi^2 v^2} (\cos 2 \beta m_Z^2 - 8 m_{\tau}^2)
    \log \left ({m_{{\tilde \tau}_1}^2  m_{{\tilde \tau}_2}^2 \over \Lambda^4} \right ) \cr
    & &\mbox{}
    + {\cos^2 2 \beta \over 16 \pi^2 v^2}
    \left ({3 m_Z^2 \over 4} - m_W^2 \right )^2 f_1 (m_{{\tilde \tau}_1}^2, \ m_{{\tilde \tau}_2}^2) \cr
    & &\mbox{}
    + {\Delta_{\tilde \tau} \over 16 \pi^2 v^2}
    (4 m_{\tau}^2 - \cos 2 \beta m_Z^2)
    {\log (m_{{\tilde \tau}_2}^2 / m_{{\tilde \tau}_1}^2)
    \over (m_{{\tilde \tau}_2}^2 - m_{{\tilde \tau}_1}^2)} \  , \cr
& & \cr
    \delta m_{\tilde \chi}^2 & = &
    {\Delta_{\tilde \chi}^2 \over 8 \pi^2 v^2}
    {f_2 (m_{{\tilde \chi}_1}^2, \ m_{{\tilde \chi}_2}^2)
    \over (m_{{\tilde \chi}_2}^2 - m_{{\tilde \chi}_1}^2)^2}
    + {m_W^4 \over 4 \pi^2 v^2}
    \log \left ({m_W^6 m_C^2 \over m_{{\tilde \chi}_1}^4 m_{{\tilde \chi}_2}^4} \right )     \cr
    & &\mbox{}
    - {\cos^2 2 \beta m_W^4 \over 2 \pi^2 v^2} f_1 (m_{{\tilde \chi}_1}^2, \ m_{{\tilde \chi}_2}^2)
    - {\Delta_{\tilde \chi} m_W^2 \over 2 \pi^2 v^2}
    {\log (m_{{\tilde  \chi}_2}^2/ m_{{\tilde  \chi}_1}^2)
    \over (m_{{\tilde  \chi}_2}^2 - m_{{\tilde  \chi}_1}^2)}   \ , \cr
& & \cr
    \delta m_h^2 & = &
    \mbox{} - {v^2 \Delta_h \over 32 \pi^2} {f_2 (m_h^2, \ m_H^2) \over (m_H^2 - m_h^2)^2}
    + {m_Z^2 \over 64 \pi^2 v^2}
    (m_A^2 \sin^2 2 \beta + 8 m_Z^2 \cos^2 \beta \cos 2 \beta) f_1 (m_h^2, \ m_H^2) \cr
    & &\mbox{}
    + {m_Z^2 \Delta_h \over 16 \pi^2} {\log (m_H^2 / m_h^2) \over (m_H^2 - m_h^2)}
    + {m_Z^4 \over 32 \pi^2 v^2} \log \left ({m_h^2 m_H^2 \over \Lambda^4} \right ) \ , \cr
    & &\mbox{}
    + {3 m_Z^4 \over 8 \pi^2 v^2} \log \left ({m_Z^2 \over \Lambda^2} \right )
    - {m_Z^4 \cos^2 2 \beta \over 128 \pi^2 v^2} (\cos 4 \beta - 5)
    \log \left ({m_A^2 \over \Lambda^2} \right )   \ , \cr
 & & \cr
    \delta m_{{\tilde \chi}^0}^2 & = &
    \sum_{k=1}^4 {m_{{\tilde \chi}^0_k}^2 \over 16 \pi^2}
    \left (\log {m_{{\tilde \chi}^0_k}^2 \over \Lambda^2} - 1 \right )
    {\displaystyle B_2 m_{{\tilde \chi}^0_k}^4 + C_2 m_{{\tilde \chi}^0_k}^2 + D_2 \over
    v^2 \prod\limits^4_{a \not= k} (m_{{\tilde \chi}^0_k}^2 - m_{{\tilde \chi}^0_a}^2)}
\cr
    & &\mbox{}
    - \sum_{k=1}^4 {1 \over 16 \pi^2}
    \log \left ({m_{{\tilde \chi}^0_k}^2 \over \Lambda^2} \right )
    {\displaystyle (A m_{{\tilde \chi}^0_k}^6 + B m_{{\tilde \chi}^0_k}^4 + C m_{{\tilde \chi}^0_k}^2 +D)^2 \over
    v^2 \left [\prod\limits^4_{a \not= k} (m_{{\tilde \chi}^0_k}^2 - m_{{\tilde \chi}^0_a}^2) \right ]^2}
\cr
    & &\mbox{}
    + \sum_{k=1}^4 {m_{{\tilde \chi}^0_k}^2 \over 8 \pi^2}
    \left (\log {m_{{\tilde \chi}^0_k}^2 \over \Lambda^2} - 1 \right )
    (A m_{{\tilde \chi}^0_k}^6 + B m_{{\tilde \chi}^0_k}^4 + C m_{{\tilde \chi}^0_k}^2 + D) \cr
    & &\mbox{}
    \times
    \left [{1\over v^2} \sum_{a \not= k}^4
    {(A m_{{\tilde \chi}^0_a}^6 + B m_{{\tilde \chi}^0_a}^4 + C m_{{\tilde \chi}^0_a}^2 + D)
    \over (m_{{\tilde \chi}^0_k}^2 - m_{{\tilde \chi}^0_a}^2)^3
    \prod\limits^4_{c \not = k \not = a} (m_{{\tilde \chi}^0_k}^2 - m_{{\tilde \chi}^0_c}^2)
    (m_{{\tilde \chi}^0_a}^2 - m_{{\tilde \chi}^0_c}^2)} \right ]
    \ ,
\end{eqnarray}
where
\begin{eqnarray}
    \Delta_{\tilde t} & = &
    2 m_t^2 (A_t^2 - 2 A_t \mu \cot \beta \cos \phi_t + \mu^2 \cot^2 \beta) \cr
& &\mbox{} + \cos 2 \beta \left ({4 m_W^2 \over 3} - {5 m_Z^2 \over
6} \right ) \left \{m_Q^2 - m_T^2 + \cos 2 \beta \left ({4 m_W^2
\over 3} - {5 m_Z^2 \over 6} \right ) \right \}  \ , \cr
    \Delta_{\tilde b} & = &
    2 m_b^2 (A_b^2 - 2 A_b \mu \tan \beta \cos \phi_b + \mu^2 \tan^2 \beta) \cr
& &\mbox{} + \cos 2 \beta \left ({m_Z^2 \over 6} - {2 m_W^2 \over 3}
\right ) \left \{m_Q^2 - m_B^2 + \cos 2 \beta \left ({m_Z^2 \over 6}
- {2 m_W^2 \over 3} \right ) \right \}  \ , \cr
    \Delta_{\tilde \tau} & = &
    2 m_{\tau}^2 (A_{\tau}^2 - 2 A_{\tau} \mu \tan \beta \cos \phi_{\tau} + \mu^2 \tan^2 \beta) \cr
& &\mbox{} + \cos 2 \beta \left ({3 m_Z^2 \over 4} - m_W^2 \right )
\left \{m_L^2 - m_E^2 + \cos 2 \beta \left ({3 m_Z^2 \over 4} -
m_W^2 \right ) \right \}  \ , \cr
    \Delta_{\tilde \chi} & = &
    2 m_W^2 (M_2^2 + \mu^2 + 2 M_2 \mu \sin 2 \beta \cos \phi_c + 2 \cos^2 2 \beta m_W^2 ) \ , \cr
    \Delta_h & = &
    {m_Z^2 \over 2 v^2} \{(3 + \cos 4 \beta) m_Z^2 - (1 + 3 \cos 4 \beta) m_A^2 \} \ ,
\nonumber
\end{eqnarray}
and
\begin{eqnarray}
    A & = &\mbox{} - 4 m_Z^2    \ ,
\cr
    B & = &\mbox{}
    4 M_1^2 m_W^2  + 4 M_2^2 (m_Z^2 - m_W^2) + 4 m_Z^2 (m_Z^2 + \mu^2)  \cr
        & &\mbox{}
    - 4 \mu \sin\beta\sin 2\beta
    \left[M_2 m_W^2 \cos \phi_2 - M_1 (m_W^2 - m_Z^2) \cos (\phi_1 + \phi_2) \right] \ ,
\cr
    C & = &\mbox{}
    - 4 m_Z^4 \mu^2 \sin^2 2 \beta -4 M_1^2 m_W^2 (m_W^2 + \mu^2)
    + 4 M_2^2 (m_Z^2 - m_W^2) (m_W^2 - m_Z^2 - \mu^2)  \cr
        & &\mbox{}
    +8 M_1 M_2 m_W^2 (m_W^2 - m_Z^2) \cos \phi_1
        + 4 M_2 m_W^2 \mu (M_1^2 + \mu^2) \sin 2 \beta \cos \phi_2    \cr
        & &\mbox{}
    + 4 M_1 \mu (m_Z^2 - m_W^2) (M_2^2 + \mu^2) \sin 2 \beta \cos (\phi_1 + \phi_2)  \ ,
\cr
    D & = &\mbox{}
    + 4 M_1^2 m_W^4 \mu^2 \sin^2 2 \beta + 4 M_2^2 \mu^2 (m_Z^2 - m_W^2)^2 \sin^2 2 \beta  \cr
        & &\mbox{}
    + 8 M_1 M_2 m_W^2 \mu^2 (m_Z^2 - m_W^2) \sin^2 2 \beta \cos \phi_1
        - 4 M_1^2 M_2 m_W^2 \mu^3 \sin 2 \beta \cos \phi_2  \cr
        & &\mbox{}
    + 4 M_1 M_2^2 \mu^3 (m_W^2 - m_Z^2) \sin 2 \beta \cos (\phi_1 + \phi_2)    \ ,
\nonumber
\end{eqnarray}
and
\begin{eqnarray}
    B_2 & = & 8 m_Z^4  \ ,
\cr
    C_2 & = &\mbox{}
    - 8 M_1^2 m_W^4 - 8 M_2^2 (m_Z^2 - m_W^2)^2 + 16 M_1 M_2 m_W^2 (m_W^2  - m_Z^2) \cos \phi_1  \ ,
\cr
    D_2 & = & 8 M_1^2 m_W^4 \mu^2 \sin^2 2  \beta + 8 M_2^2  \mu^2 (m_Z^2 - m_W^2)^2 \sin^2 2 \beta  \cr
    & &\mbox{}
    + 16 M_1 M_2 m_W^2 \mu^2 (m_Z^2 - m_W^2) \sin^2 2 \beta \cos \phi_1  \ .
\nonumber
\end{eqnarray}
Here the scale independent function $f_2 (m_x^2, m_y^2)$ is defined by
\[
    f_2 (m_x^2, m_y^2) =
    {m_y^2 + m_x^2 \over m_y^2 - m_x^2} \log {m_y^2 \over m_x^2} - 2 \,.\]
Note that the above upper bound on the lightest neutral Higgs boson mass
does not depend on the tree-level mass of the pseudoscalar Higgs boson.

It is also possible to express analytic forms for the squared masses of the three Higgs bosons as
\begin{equation}
m_{h_n}^2 = {1 \over 3} {\rm Tr}(M_h) + 2 \sqrt{W} \cos \left \{ {
\theta + 2 n \pi \over 3 } \right \} \ , ~(n=1,2,3)
\end{equation}
with
\begin{equation}
\theta = \cos^{-1} \left ({Y \over \sqrt{W^3}} \right ) \ ,
\end{equation}
where
\begin{eqnarray}
W & = &\mbox{} - {1 \over 18} \{{\rm Tr}(M_h) \}^2 + {1 \over 6}
{\rm Tr} (M_h^2) \ , \cr Y & = &\mbox{} - {5 \over 108} \{ {\rm Tr}
(M_h) \}^3 + {1 \over 12} {\rm Tr} (M_h) {\rm Tr} (M_h^2) + {1 \over
2} {\rm det} (M_h) \ .
\end{eqnarray}
One can calculate the transformation matrix for the neutral Higgs bosons from the orthogonality condition.
The elements of the transformation matrix are given by
\begin{equation}
U_{ij} = {o_{ij} \over \sqrt{\sum_{k=1}^3 o_{ik}^2} } \ ,
\end{equation}
where
\begin{eqnarray}
o_{ii} & = & 1  \ , ~(i=1,2,3), \cr o_{12} & = & {(m_{h_1}^2 -
M_{11}) M_{23} + M_{12} M_{13} \over (m_{h_1}^2 - M_{22}) M_{13} +
M_{12} M_{23} } \ ,  \cr o_{13} & = & {(m_{h_1}^2 - M_{11})
(m_{h_1}^2 - M_{22}) - M_{12}^2 \over (m_{h_1}^2 - M_{22}) M_{13} +
M_{12} M_{23} } \ , \cr o_{21} & = & {(m_{h_2}^2 - M_{22}) M_{13} +
M_{12} M_{23} \over (m_{h_2}^2 - M_{11}) M_{23} + M_{12} M_{13} } \
,  \cr o_{23} & = & {(m_{h_2}^2 - M_{11}) (m_{h_2}^2 - M_{22}) -
M_{12}^2 \over (m_{h_2}^2 - M_{22}) M_{23} + M_{12} M_{13} } \ , \cr
o_{31} & = & {(m_{h_3}^2 - M_{33}) M_{12} + M_{23} M_{13} \over
(m_{h_3}^2 - M_{11}) M_{23} + M_{12} M_{13} } \ ,  \cr o_{32} & = &
{(m_{h_3}^2 - M_{11}) (m_{h_3}^2 - M_{33}) - M_{13}^2 \over
(m_{h_3}^2 - M_{11}) M_{23} + M_{12} M_{13} } \ .
\end{eqnarray}

\section{Higgs productions}

The most important channels for the productions of neutral Higgs bosons at the LHC are:
the gluon fusion process via the triangular loop of top quark $PP \rightarrow gg \rightarrow h_i$,
the Higgs-strahlung process mediated by $W$ boson $PP \rightarrow q {\bar q'} \rightarrow W h_i$,
and the Higgs-strahlung process mediated by $Z$ boson $PP \rightarrow q {\bar q} \rightarrow Z h_i$.
We denote the production cross sections for these processes as $\sigma (h_i)$, $\sigma (W h_i)$,
and $\sigma (Z h_i)$ respectively.
At the lowest order, the cross sections of these processes are related to the SM cross sections
for the corresponding SM Higgs boson production channels as [44-47]
\begin{eqnarray}
    & & \sigma (h_i) = K_i^2 \sigma_{\rm SM} (PP \rightarrow gg \rightarrow h_i)  \ , \cr
    & & \sigma (W h_i) = R_i^2 \sigma_{\rm SM} (PP \rightarrow q {\bar q'} \rightarrow W h_i) \ ,  \cr
    & & \sigma (Z h_i) = R_i^2 \sigma_{\rm SM} (PP \rightarrow q {\bar q} \rightarrow Z h_i) \ ,
\end{eqnarray}
where $K_i$ and $R_i$ ($i$ = 1,2,3) are defined as
\begin{eqnarray}
    K_i & = & {U_{2i} \over \sin \beta} \ , \cr
    R_i & = & \cos \beta U_{1i} + \sin \beta U_{2i} \ ,
\end{eqnarray}
with $U_{ij}$ being the elements of the diagonalization matrix
for the $3 \times 3$ neutral Higgs boson mass matrix.
Using the ortho-normality condition of $U_{ij}$, one can show that $\sum_{i = 1}^3 R_i^2 = 1$.
The factor $K_i$ comes from the coupling of the $i$th neutral Higgs boson
to a top quark pair normalized by the corresponding SM coupling, and
the factor $R_i$ comes from the coupling of the $i$th neutral Higgs boson
to a $Z$ ($W$) boson pair normalized by the corresponding SM coupling.

For the numerical analysis, we take $\sin^2 \theta_W = 0.23$ for weak-mixing angle,
$G_F = 1.166 \times 10^{- 5}$ for Fermi coupling constant,
$m_Z = 91.187$ GeV for the $Z$ boson mass,
$m_W = 80.423$ GeV for the $W$ boson mass,
$m_t = 175$ GeV for the mass of the top quark,
and $m_b = 4.5$ GeV for the mass of the bottom quark.
The renormalization and factorization scales are taken to be the same
as the neutral Higgs boson mass.
The parton densities are set as CTEQ6M [48,49].

We would like to concentrate on a particular region of the parameter space of the MSSM
with explicit CP violation.
The relevant free parameters are $\Lambda$, $\tan \beta$, $\mu$, $m_Q$, $m_T$,
$m_B$, $m_L$, $m_E$, $A_t$, $A_b$, $A_{\tau}$, $M_1$, $M_2$,
$\phi_t$, $\phi_b$, $\phi_{\tau}$, $\phi_c (\phi_2)$, and $\phi_1$.
We set $m_Q = m_L$, $m_T = m_B = m_E$,
$A_t = A_b = A_{\tau}$, and $\phi_t = \phi_b = \phi_{\tau} = \phi_c (\phi_2)$ for simplicity.
At the electroweak scale one can take the relation between $U(1)$ and $SU(2)$ gaugino masses
to be $M_1 = 5 \tan^2 \theta_W M_2/3$.

We are interested in examining the dependence of the contribution of the neutralinos
on the production cross section of the neutral Higgs bosons.
The contributions of the neutralino loops depend crucially on the CP phase $\phi_1$.
In other words, the CP phase $\phi_1$ occurs only in the expressions from the neutralino contributions.
We search for some parameter region where the three neutral Higgs bosons becomes
relatively light with masses below 150 GeV by using a random number generating function.
In such a region, the production cross sections of the neutral Higgs bosons mainly depend
on the relevant coupling constants rather than the neutral Higgs boson masses.
We calculate the neutral Higgs boson masses as $\phi_1$ varies from zero to $\pi$, for
$\tan \beta = 28.2$, ${\bar m}_A$ = 135 GeV,
$\mu= 458$ GeV, $m_Q$ = 544 GeV, $m_T$ = 480 GeV, $A_t$ = 932 GeV,
$M_2$ = 390 GeV, and $\phi_t$ = 2.58.
In this case, the neutral Higgs boson masses depend weakly on the CP phase $\phi_1$
and are given approximately by $m_{h_1}\approx$ 131 GeV,
$m_{h_2}\approx$ 132 GeV, and $m_{h_3}\approx$ 135 GeV.

We calculate the value of $K_i^2$ and $R_i^2$ as functions of the CP phase $\phi_1$, for the above parameter values.
These values are plotted in Fig.1 as the phase $\phi_1$ varies from zero to $\pi$.
Note that the phase dependence of $K_i^2$ is carried by $U_{2i}$ via mass matrix elements
and that of $R_i^2$ is carried by both $U_{1i}$ and $U_{2_i}$.
The production cross sections of the neutral Higgs bosons via the gluon fusion process mediated
by the triangular loop of the top quark are modified by $K_i$ from the corresponding SM cross section,
and the production cross sections of the neutral Higgs bosons via the Higgs-strahlung process mediated
by the vector bosons are modified by $R_i^2$ from the corresponding SM cross section.
One can find from Fig. 1 that the factors vary 12 \% for $K_1^2$, 19 \% for $K_2^2$, and 3.6 \% for $K_3^2$,
while the factors vary 13 \% for $R_1^2$, 18 \% for $R_2^2$, and 4.2 \% for $R_3^2$.

It is well known that the gluon fusion process at the LHC is the most dominant mechanism for the production
of the neutral Higgs boson in the SM.
It is also well known that the Higgs-strahlung process mediated by the $W$ boson at the LHC is more dominant
than that mediated by the $Z$ boson for the production of the neutral Higgs boson in the SM.

In Fig. 2, the production cross sections of th neutral Higgs bosons for the processes in Eq. (20) at the LHC are
plotted as functions of the CP phase $\phi_1$.
The values of the other parameters are taken to be the same as in Fig. 1.

In this figure, the solid curves represent the production cross sections of the neutral Higgs bosons
via the gluon fusion process which is mediated with the triangular loop of the top quark.
One can see that $\sigma (h_1)$, $\sigma (h_2)$, and $\sigma (h_3)$ vary
about 13.0 \%, 19.5 \%, and 3.8 \% respectively as $\phi_1$ goes from zero to $\pi$.
The production cross section of the $i$th neutral Higgs boson via the gluon fusion depends significantly on the factors $K_i$.

The dashed curves represent the production cross sections of the neutral Higgs bosons
via the Higgs-strahlung process mediated by the $W$ boson.
One can see that $\sigma (W h_1)$, $\sigma (W h_2)$, and $\sigma (W h_3)$ vary
about 14.1 \%, 18.3 \%, and 3.8 \% respectively as $\phi_1$ goes from zero to $\pi$.
The dotted curves represent the production cross sections of the neutral Higgs bosons via the Higgs-strahlung process
mediated by the $Z$ boson.
One can also see that $\sigma (Z h_1)$, $\sigma (Z h_2)$, and $\sigma (Z h_3)$ vary
about 14.1 \%, 12.7 \%, and 3.4 \% respectively as $\phi_1$ goes from zero to $\pi$.
The production cross sections of the $i$th neutral Higgs boson via the the Higgs-strahlung processes
mediated by the vector bosons depend significantly on the factors $R_i$.
Note that the variations of the above cross sections are different from those of the factors $K_i$ and $R_i$ as can be seen by comparing Figs. 1 and 2.
This is because the corresponding SM cross sections also depend on the phase $\phi_1$ indirectly through the masses of the neutral Higgs bosons.

We now turn to the production processes of the Higgs boson accessible at the ILC.
At the ILC, the dominant production mechanisms for the neutral Higgs boson are the Higgs-strahlung
process and the vector boson ($W$, $Z$) fusion processes.
At the lowest order, the cross sections of these processes are related to the SM cross sections
for the corresponding SM Higgs boson production processes as [50-53]
\begin{eqnarray}
    & & \sigma (ZZ h_i) = R_i^2 \sigma_{\rm SM} (e^+e^- \rightarrow Z h_i)  \ , \cr
    & & \sigma (\nu \nu h_i) = R_i^2 \sigma_{\rm SM} (e^+e^- \rightarrow \nu_e {\bar \nu}_e h_i) \ ,  \cr
    & & \sigma (ee h_i) = R_i^2 \sigma_{\rm SM} (e^+e^- \rightarrow e^+e^- h_i) \ .
\end{eqnarray}

It is well known that at the center of mass energy of $\sqrt{s}\sim 200$ GeV,
the Higgs-strahlung process is the most dominant mechanism for the production of the neutral Higgs boson in the SM.
It is also well known that at the center of mass energy of $\sqrt{s}\sim 500$ GeV,
the $WW$ fusion process is the most dominant mechanism for the production of the neutral Higgs boson in the SM.

In Fig. 3, the production cross sections of the neutral Higgs bosons for the processes in Eq. (22)
at the ILC with $\sqrt{s} = 500$ GeV (ILC500) are plotted as functions of the CP phase $\phi_1$.
The values of the other parameters are taken to be the same as in Fig. 1.

In this figure, the solid curves represent the production cross sections of the neutral Higgs bosons
via the Higgs-strahlung process at the ILC500.
One can see that $\sigma (ZZ h_1)$, $\sigma (ZZ h_2)$, and $\sigma (ZZ h_3)$
vary about 14.1 \%, 18.3 \%, and 4 \% respectively as $\phi_1$ goes from zero to $\pi$.

The dashed curve represent the production cross sections of the neutral Higgs bosons via the $WW$ fusion process.
One can see that $\sigma (\nu \nu h_1)$, $\sigma (\nu \nu h_2)$, and $\sigma (\nu \nu h_3)$
vary about 14.1 \%, 18.3 \%, and 4 \% respectively as $\phi_1$ goes from zero to $\pi$.

The dotted curve represent the production cross sections of the neutral Higgs bosons via the $ZZ$ fusion process.
One can also see that $\sigma (ee h_1)$, $\sigma (ee h_2)$, and $\sigma (ee h_3)$ vary
about 14.1 \%, 16.7 \%, and 3.8 \% respectively as $\phi_1$ goes from zero to $\pi$.
The production cross sections of the $i$th neutral Higgs bosons via the three dominant processes at the ILC500
depend significantly on the factors $R_i$.

Here again, one can see that the variations of the above cross sections are different from those of the factor $R_i$
as can be seen by comparing Figs. 1 and 3.
It is interesting to notice that the variations for the solid curves are almost the same as those for the dashed curves,
which indicates that the dependences of $\sigma_{\rm SM} (e^+e^- \rightarrow Z h_i)$
and $\sigma_{\rm SM} (e^+e^- \rightarrow \nu_e {\bar \nu}_e h_i)$ on the CP phase $\phi_1$ are almost identical.

\section{Conclusions}

We have studied the Higgs sector of the MSSM where the CP symmetry
is explicitly violated at the one-loop level.
We have calculated the mass matrix of the three neutral Higgs bosons and
the analytic forms of their masses.
In this calculation, we have taken into account all the relevant one-loop contributions
including those of the top quark, the stop quarks, the bottom quark, the sbottom quarks,
the tau lepton, the stau leptons, the $W$ boson, the charged Higgs boson,
the charginos, the $Z$ boson, the neutral Higgs bosons, and the neutralinos.

We have also studied the production processes of the neutral Higgs bosons.
Three dominant channels accessible at the LHC for the neutral Higgs boson production
are the Higgs-strahlung processes mediated by the $Z$ boson, the $W$ boson,
and the gluon fusion process involving the triangular loop of the top quark,
and three dominant channels accessible at the ILC for the neutral Higgs boson production
are the Higgs-strahlung process, $WW$ and $ZZ$ fusion processes.
We have calculated the cross sections of these processes.
The range of the variations of the cross sections due to the variation of the CP phase $\phi_1$
turns out to be as high as about 19.5 \%.
This indicates the nontrivial dependence of the Higgs production processes
on the CP phase $\phi_1$ which arises from the neutralino sector.

In the MSSM with explicit CP violation, radiative corrections of the neutralino loop
to the tree-level Higgs sector give rise to an important contribution
for the CP mixing between the scalar and pseudoscalar Higgs bosons.
Therefore, we suggest that, as in the explicit CP violation scenario,
such a CP mixing effect arising from the neutralino contribution might have
important phenomenological implications for the Higgs search at both the LHC and the ILC.

\vskip 0.3 in
\noindent
{\large {\bf ACKNOWLEDGMENTS}}
\vskip 0.2 in
This work is partly supported by KOSEF through a grant provided by the MOST in 2007
(project No. K2071200000107 A020000110) and by KOSEF through CHEP, Kyungpook National University.

\vfil\eject


\vfil\eject

{\large {\bf FIGURE CAPTION}}
\vskip 0.3 in
\noindent
FIG. 1. : The factors $K_i^2$ and $R_i^2$ ($i=1,2,3$) as functions of $\phi_1$,
for $\tan \beta = 28.2$, ${\bar m}_A = 135$ GeV, $\mu= 458$ GeV, $m_Q
= m_L = 544$ GeV, $m_T = 480$ GeV, $A_t = 932$ GeV, $M_2 = 390$ GeV,
and $\phi_t  = 2.58$.

\vskip 0.3 in
\noindent
FIG. 2. : The cross sections for $h_i$ ($i=1,2,3$) productions
as functions of $\phi_1$, via the gluon fusion process mediated by the
triangular loop of the top quark (solid curves),
via the Higgs-strahlung processes mediated by the $W$ boson (dashed curves) and the $Z$ boson (dotted curves).
The other parameters are the same as in Fig. 1.

\vskip 0.3 in
\noindent
FIG. 3. : The cross sections for $h_i$ ($i=1,2,3$) productions
as functions of $\phi_1$, via the Higgs-strahlung process (solid curves),
the $WW$ fusion process (dashed curves), and the $WW$ fusion process (dotted curves) at the ILC with $\sqrt{s} = 500$ GeV.
The other parameters are the same as in Fig. 1.

\vfil\eject
\setcounter{figure}{0}
\def\figurename{}{}%
\renewcommand\thefigure{FIG. 1}
\begin{figure}[t]
\begin{center}
\includegraphics[scale=0.6]{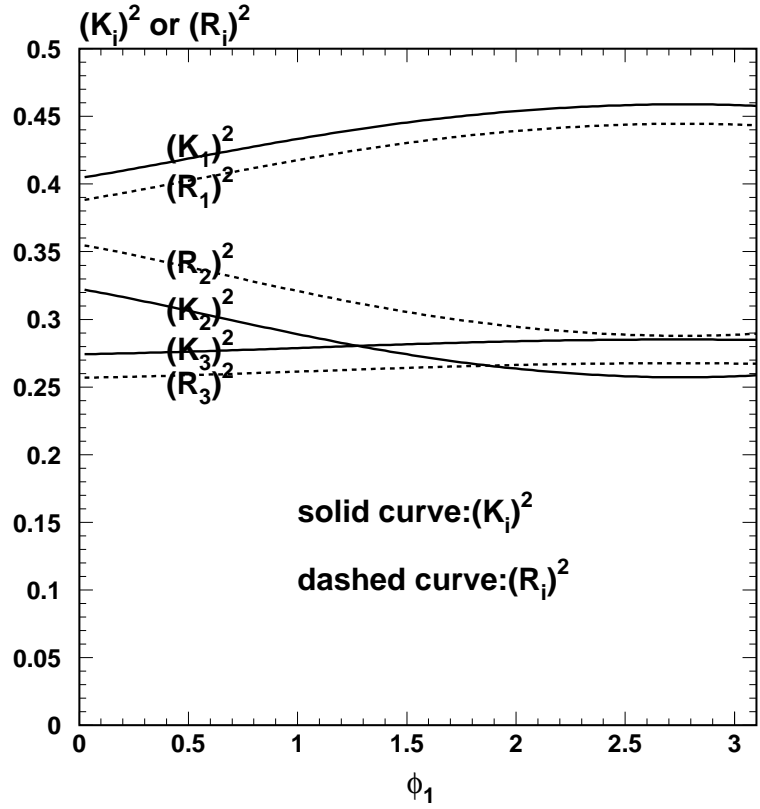}
\caption[plot]{The factors $K_i^2$ and $R_i^2$ ($i=1,2,3$) as functions of $\phi_1$,
for $\tan \beta = 28.2$, ${\bar m}_A = 135$ GeV, $\mu= 458$ GeV, $m_Q
= m_L = 544$ GeV, $m_T = 480$ GeV, $A_t = 932$ GeV, $M_2 = 390$ GeV,
and $\phi_t  = 2.58$.}
\end{center}
\end{figure}

\renewcommand\thefigure{FIG. 2}
\begin{figure}[t]
\begin{center}
\includegraphics[scale=0.6]{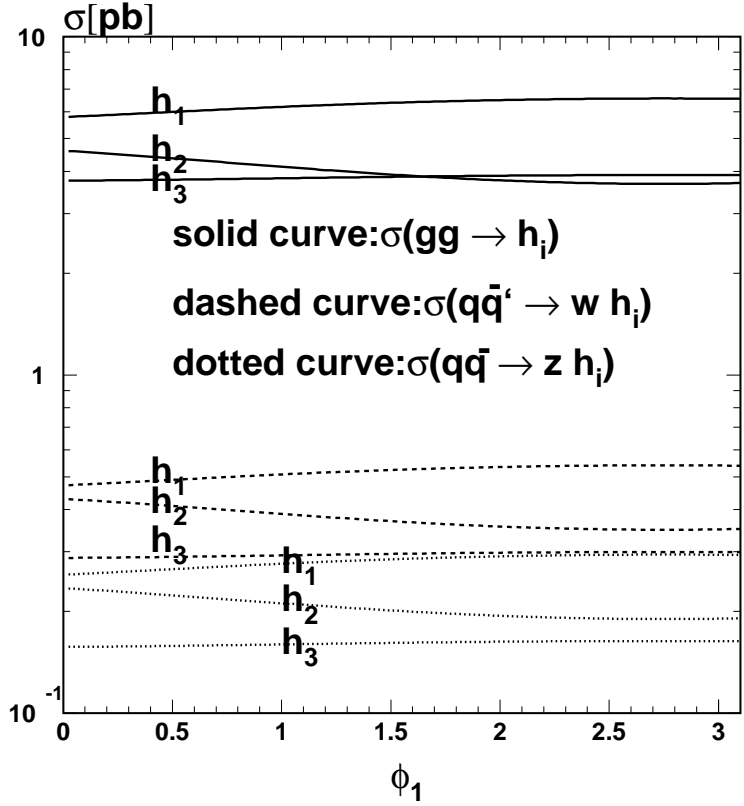}
\caption[plot]{The cross sections for $h_i$ ($i=1,2,3$) productions
as functions of $\phi_1$, via the gluon fusion process mediated by the
triangular loop of the top quark (solid curves),
via the Higgs-strahlung processes mediated by the $W$ boson (dashed curves) and the $Z$ boson (dotted curves).
The other parameters are the same as in Fig. 1.}
\end{center}
\end{figure}

\renewcommand\thefigure{FIG. 3}
\begin{figure}[t]
\begin{center}
\includegraphics[scale=0.6]{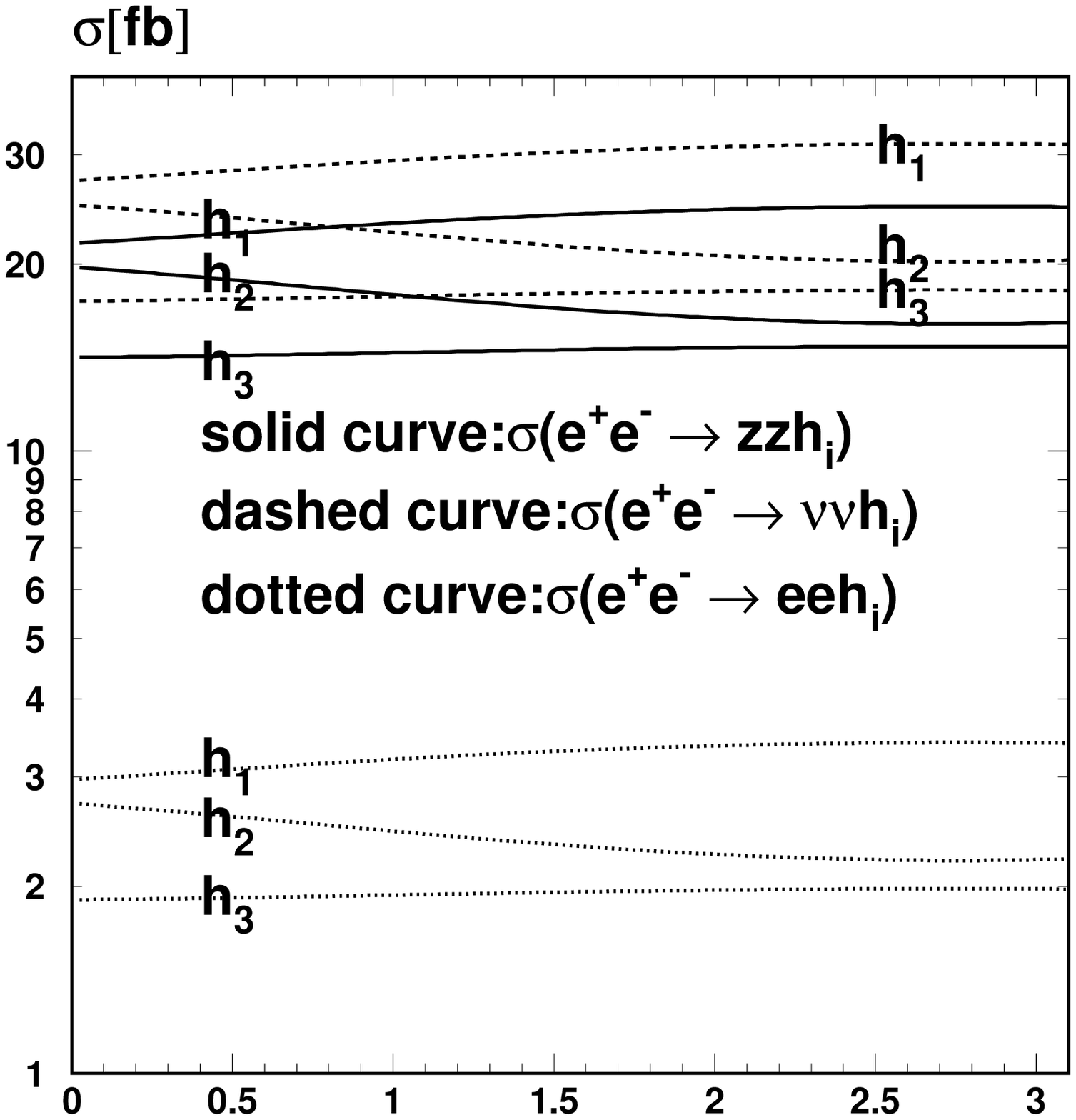}
\caption[plot]{The cross sections for $h_i$ ($i=1,2,3$) productions
as functions of $\phi_1$, via the Higgs-strahlung process (solid curves),
the $WW$ fusion process (dashed curves), and the $WW$ fusion process (dotted curves) at the ILC with $\sqrt{s} = 500$ GeV.
The other parameters are the same as in Fig. 1.}
\end{center}
\end{figure}

\begin{thebibliography}{99}
\bibitem{1} J. H. Christen, J. W. Cronin, V. L. Fitch, and R. Turlay, Phys. Rev. Lett. {\bf 13}, 138 (1964).
\bibitem{2} N. Cabibbo, Phys. Rev. Lett. {\bf 10}, 531 (1963);
M. Kobayashi and T. Maskawa, Prog. Theor. Phys. {\bf 49}, 652
(1973).
\bibitem{3} S. Weinberg, Phys. Rev. Lett. {\bf 37}, 657 (1976).
\bibitem{4} P. Fayet and S. Ferrara, Phys. Rep. {\bf 32}, 249 (1977);
    P. Fayet, Phys. Rep. {\bf 105}, 21 (1984).
\bibitem{5} H. P. Nilles, Phys. Rep. {\bf 110}, 1 (1984).
\bibitem{6} H. E. Haber and G. L. Kane, Phys. Rep. {\bf 117}, 75 (1985).
\bibitem{7} J. F. Gunion, H. E. Haber, G. L. Kane, and S. Dawson, {\it The Higgs Hunters' Guide}
(Addison-Wesley, Redwood City, CA, USA, 1990).
\bibitem{8} N. Maekawa, Phys. Lett. B {\bf 282}, 387 (1992).
\bibitem{9} A. Pomarol, Phys. Lett. B {\bf 287}, 331 (1992).
\bibitem{10} N. Haba, Phys. Lett. B {\bf 398}, 305 (1997).
\bibitem{11} O. Lebedev, Mod. Phys. Lett. A {\bf 13}, 735 (1998).
\bibitem{12} O. Lebedev, Eur. Phys. J. C {\bf 4}, 363 (1998).
\bibitem{13} M. Brhlik and G. L. Kane, Phys. Lett. B {\bf 437}, 331 (1998).
\bibitem{14} A. Pilaftsis and C. E. M. Wagner, Nucl. Phys. B {\bf 553}, 3 (1999).
\bibitem{15} D. A. Demir, Phys. Rev. D {\bf 60}, 095007 (1999).
\bibitem{16} S. Y. Choi and J. S. Lee, Phys. Rev. D {\bf 61}, 015003 (1999).
\bibitem{17} M. Carena, J. Ellis, A. Pilaftsis, and C. E. M. Wagner, Nucl. Phys. B {\bf 586}, 92 (2000).
\bibitem{18} G. L. Kane and L. T. Wang, Phys. Lett. B {\bf 488}, 383 (2000).
\bibitem{19} T. Ibrahim and P. Nath, Phys. Rev. D {\bf 63}, 035009 (2001);
Phys. Rev. D {\bf 66}, 015005 (2002); Phys. Rev. D {\bf 68}, 015008
(2003); Phys. Rev. D {\bf 70}, 035006 (2004).
\bibitem{20} S. Heinemeyer, Eur. Phys. J. C {\bf 22}, 521 (2001).
\bibitem{21} M. Boz and N. K. Pak, Phys. Rev. D {\bf 65}, 075014 (2002).
\bibitem{22} M. Carena, J. Ellis, A. Pilaftsis, and C. E. M. Wagner, Nucl. Phys. B {\bf 625}, 345 (2002).
\bibitem{23} E. Christova, H. Eberl, S. Kraml, and W. Majerotto, Nucl. Phys. B {\bf 639}, 263 (2002).
\bibitem{24} N. Ghodbane, S. Katsanevas, I. Laktineh, J. Rosiek, Nucl. Phys. B {\bf 647}, 190 (2002).
\bibitem{25} M. Boz,  J. Phys. G {\bf 28}, 2377 (2002).
\bibitem{26} S. W. Ham, S. K. Oh, E. J. Yoo, and H. K. Lee, J. Phys. G {\bf 27}, 1 (2001).
\bibitem{27} S. W. Ham, S. K. Oh, E. J. Yoo, C. M. Kim, and D. Son, Phys. Rev. D {\bf 68}, 055003 (2003).
\bibitem{28} M. Carena and H. E. Haber, Prog. Part. Nucl. Phys. {\bf 50}, 63 (2003).
\bibitem{29} A. Dedes and A. Pilaftsis, Phys. Rev. D {\bf 67}, 015012 (2003).
\bibitem{30} M. Carena, J. R. Ellis, S. Mrenna, A. Pilaftsis, and C. E. M. Wagner, Nucl. Phys. B {\bf 659}, 145 (2003).
\bibitem{31} J. S. Lee, A. Pilaftsis, M. Carena, S. Y. Choi, M. Drees, J. R. Ellis, and C. E. M. Wagner,
Comput. Phys. Commun. {\bf 156}, 283 (2004).
\bibitem{32} J.R. Ellis, J. S. Lee, and A. Pilaftsis, Phys. Rev.D {\bf 70}, 075010, (2004).
\bibitem{33} M. Argyrou, A. B. Lahanas, and D. V. Nanopoulos, Phys. Rev. D {\bf 70}, 095008, (2004); Erratum-ibid.D {\bf 70}, 119902, (2004).
\bibitem{34} A.G. Akeroyd and A. Arhrib, Phys. Rev. D {\bf 64} 095018 (2001).
\bibitem{35} M. Carena, J. Ellis, A. Pilaftsis, and C. E. M. Wagner, Phys. Lett. B {\bf 495}, 155 (2000).
\bibitem{36} A. Arhrib, Phys. Rev. D {\bf 67}, 015003 (2003).
\bibitem{37} A. Arhrib, D.K. Ghosh, and Otto C.W. Kong, Phys. Lett. B {\bf 537}, 217 (2002).
\bibitem{38} S.Y. Choi, K. Hagiwara, and J.S. Lee, Phys. Lett. B {\bf 529}, 212 (2002).
\bibitem{39} A. Mendez and A. Pomarol, Phys. Lett. B {\bf 272}, 313 (1991).
\bibitem{40} J. F. Gunion, B. Grzadkowski, H. E. Haber, and J. Kalinowski, Phys. Rev. Lett. {\bf 79}, 982 (1997).
\bibitem{41} B. Grzadkowski, J. F. Gunion, and J. Kalinowski, Phys. Rev. D {\bf 60}, 075001 (2000);
B. Grzadkowski, J. F. Gunion, and J. Kalinowski, Phys. Lett. B {\bf
480}, 287 (2000).
\bibitem{42} S. Coleman and E. Weinberg, Phys. Rev. D {\bf 7}, 1888 (1973).
\bibitem{43} M. Drees, Int. J. Mod. Phys. A {\bf 4}, 3635 (1989).
\bibitem{44} H.M. Georgi, S.L. Glashow, M.E. Machacek, D.V. Nanopoulos, Phys. Rev. Lett. {\bf 40}, 692 (1978).
\bibitem{45} M. Spira, A. Djouadi, D. Graudenz, P.M. Zerwas, Nucl. Phys. B {\bf 453}, 17 (1995).
\bibitem{46} M. Spira, Fortsch. Phys. {\bf 46}, 203 (1998).
\bibitem{47} R.V. Harlander, Phys. Lett. B {\bf 492}, 74 (2000).
\bibitem{48} H.L. Lai, J. Botts, J. Huston, J.G. Morfin, J.F. Owens, J.W. Qiu, W.K. Tung, H. Weerts,
Phys. Rev. D {\bf 51}, 4763 (1995).
\bibitem{49} J. Pumplin, D.R. Stump, J. Huston, H.L. Lai, P. Nadolsky, and W.K. Tung, JHEP 0207, 012 (2002);
D. Stump, J. Huston, J. Pumplin, W.K. Tung, H.L. Lai, S. Kuhlmann,
and J.F. Owens, JHEP 0310, 046 (2003).
\bibitem{50} E. Accomando {\it et. al.}, Phys. Rep. {\bf 299}, 1 (1998).
\bibitem{51} A. Djouadi, PM/98-17, GDR-S-012 (1998).
\bibitem{52} T. Abe {\it et al.}, American Linear Collider Working Group, hep-ex/0106056.
\bibitem{53} TESLA, Part III: Physics at an $e^+ e^-$ Linear Collider,
edited by R. D. Heuer, D. Miller, F. Richard, and P. M. Zerwas,
hep-ph/0106315.
\end{thebibliography}
\end{document}